\documentclass{article}

\usepackage{arxiv}

\usepackage[utf8]{inputenc} 
\usepackage[T1]{fontenc}    
\usepackage{hyperref}       
\usepackage{url}            
\usepackage{booktabs}       
\usepackage{amsfonts}       
\usepackage{nicefrac}       
\usepackage{microtype}      
\usepackage{lipsum}		
\usepackage{graphicx}
\usepackage{natbib}
\usepackage{doi}
\usepackage{tabulary}

\title{Cross-domain Voice Activity Detection with Self-Supervised Representations}

\date{} 					

\author{ 
    Sina Alisamir \\
	\textit{Atos \& LIG}\\
	\textit{Univ. Grenoble Alpes}\\
	Grenoble, France \\
	sina.alisamir@univ-grenoble-alpes.fr \\
	\And
	Fabien Ringeval \\
	\textit{Grenoble INP, LIG}\\
	\textit{Univ. Grenoble Alpes, Inria, CNRS}\\
	Grenoble, France \\
	fabien.ringeval@univ-grenoble-alpes.fr \\
	\And
	François Portet \\
	\textit{Grenoble INP, LIG}\\
	\textit{Univ. Grenoble Alpes, Inria, CNRS}\\
	Grenoble, France \\
	francois.portet@univ-grenoble-alpes.fr \\
}

\renewcommand{\headeright}{}
\renewcommand{\undertitle}{}
\renewcommand{\shorttitle}{Cross-domain Voice Activity Detection with Self-Supervised Representations}

\hypersetup{
pdftitle={Voice Activity Detection with Self-Supervised Representations},
pdfsubject={},
pdfauthor={},
pdfkeywords={},
}

\begin{document}
\maketitle

\begin{abstract}
Voice Activity Detection (VAD) aims at detecting speech segments on an audio signal, which is a necessary first step for many today's speech based applications. Current state-of-the-art methods focus on training a neural network exploiting features directly contained in the acoustics, such as Mel Filter Banks (MFBs). Such methods therefore require an extra normalisation step to adapt to a new domain where the acoustics is impacted, which can be simply due to a change of speaker, microphone, or environment. In addition, this normalisation step is usually a rather rudimentary method that has certain limitations, such as being highly susceptible to the amount of data available for the new domain. Here, we exploited the crowd-sourced Common Voice (CV) corpus to show that representations based on Self-Supervised Learning (SSL) can adapt well to different domains, because they are computed with contextualised representations of speech across multiple domains. SSL representations also achieve better results than systems based on hand-crafted representations (MFBs), and off-the-shelf VADs, with significant improvement in cross-domain settings.
\end{abstract}


\section{Introduction}

The very first step to most of speech based applications such as Automatic Speech Recognition (ASR) or Speaker Emotion Recognition (SER), is Voice Activity Detection (VAD)~\cite{eyben2015real, cen2016real, harar2017speech, tong2014evaluating}. Indeed, in any speech based system, we must first identify and isolate speech segments coming from the continuous input audio stream, which may include environmental noises. 

Current VAD systems rely on supervised deep learning models trained on acoustic features such as Mel Filter Banks (MFBs), to detect speech segments. Subsequently, in real life usage, such models are usually confronted with a domain mismatch issue, because the data, as represented by the MFBs, may differ significantly from those used for model training. This problem is usually tackled by a normalisation of the acoustic features with statistics computed on the novel data~\cite{Eyben13-RVA}, but this approach is rather rudimentary and further requires to have a sufficient amount of samples of the new domain to be effective. 


On the other hand, Self-Supervised Learning (SSL) methods, such as Wav2Vec2~\cite{baevski2020wav2vec}, have shown improvements in different speech related tasks when used as speech representations compared to traditional acoustic features~\cite{evain2021task, keesing2021acoustic}. Indeed, speech is not directly represented by its spectral information, but rather by generic descriptors that aim to predict whether the acoustic is likely to be coming from distractors or genuine speech samples. This makes SSL representations more robust against domain mismatch issues~\cite{latif2020deep, alisamir2021evolution}. Training of the model can also be further realised on several different domains to improve robustness against unseen data~\cite{hsu2021robust}.


Given that SSL representations have improved many speech downstream tasks, here, we investigate their relevance for the task of VAD on a large variety of speech data collected with close-talk microphones at home, with the ultimate goal of driving a virtual assistant at home in the context of digital therapies~\cite{Tarpin-Bernard21-TDT}. We compare the results of Wav2Vec2 (W2V2) representations to traditional MFB features and off-the-shelf VADs. We also study the impact of real-time processing where less \textit{a priori} knowledge about the target domain is available to perform VAD.




\section{Related Work}
\label{sec:related}


Many current state-of-the-art VAD system mainly focus on different deep learning architectures that take acoustic features as input and then are tested on the same data distribution as the training partition \cite{jung2018joint, lee2020dual, zheng2020mlnet}. Whereas this approach can perform well on controlled experimental data, it fails to achieve similar performance on unseen data, which is problematic for real life use cases. 

Furthermore, having mostly ASR in mind, most use cases of VADs today are to first simply detect non silent segments and then use a noise robust Speech-To-Text (STT) model to predict transcriptions. In the end, speech segments that do not contain any verbal information that can be detected by the model are discarded \cite{yoshimura2020end}, which is problematic for speech technologies relying on the non-verbal analysis of speech~\cite{Tarpin-Bernard21-TDT}.

Recurrent Neural Networks, as equipped with Long Short-Term Memory cells (LSTM-RNN), have shown better generalisation capabilities and overall performance compared to state-of-the-art methods based on statistical algorithms~\cite{eyben2015real}. Gated Recurrent Units (GRU), which are a simplified version of the LSTM cells, have also been investigated in combination with CNNs, and showed improvement over state-of-the-art models
~\cite{wang2019rnn}. However,  
all studies reported degraded performance when testing on domains different from those used for model learning, 
showing that VAD systems based on deep learning architectures fail to generalise to new domains when represented by acoustic features of the speech signal. 


Unsupervised learning techniques alleviate the need for labels. They allow to learn how data are generated by exposing a model to huge amounts of samples where predictions related to it are learnt. SSL representations of speech have recently been shown to be significantly more robust against domain mismatch, compared to using acoustic features~\cite{gimeno2021unsupervised}, on conversational audio from Apollo space missions~\cite{Joglekar2021Fearless}.

\section{Method}
\label{sec:method}



\subsection{Dataset}
As our ultimate objective is to drive a virtual assistant supporting digital therapies at the home of patients, we focused on detecting French speech in close-talking microphones collected in home scenarios, and used the French partition (cv-corpus-7.0-2021-07-21) of the Common Voice (CV) corpus, which is a gigantic collection of transcribed read speech collected by the Mozilla foundation with crowd-sourcing techniques~\cite{ardila2019common}. Each speaker reads a series of text from Wikipedia at home using their own recording devices. 
Overall, there are more than 410k utterances collected from 600 speakers, and for a total duration of 16.42 hours.
The CV corpus is therefore highly relevant to investigate the robustness of VAD systems over multiple domains at once. 

\subsection{Data pre-processing}
The CV corpus is provided with segmented speech utterances. Since models based on LSTM or GRU are able to exploit long-term dependencies between input and output data, they may preferably be provided with long sequences of audio signals. Speech utterances of the Common Voice dataset were thus concatenated and interspersed with a pause of randomly chosen duration. The distribution of pause duration as observed in real conversations~\cite{ringeval2013introducing} was approximated with a Gaussian distribution; $\mu = 2.22$ seconds, $\sigma = 1.83$ seconds. 
Because speech utterances are further combined with four different types of noises, and three different levels of noise, we did not use all the data of the CV corpus, and exploited a portion of it summing up to an overall duration of at least 2 hours per speaker-independent partitions; the length of each partition is summarised in Table \ref{tab:dataset}.

\subsection{Synthetic noise degradation}
Since most of the recorded utterances present in the CV corpus have little to no noise, and that our objective is to detect voice specially in the presence of home noise, we first need to gather noise data before adding them to speech utterances. 
\subsubsection{Gathering noise data}

There exists several data sets of acoustic noises, such as DCASE or NIGENS~\cite{trowitzsch2019nigens}, but as their focus is on general sound event detection, we gathered our own noise data set, focusing on noises that are more likely to be present in the acoustic background of an average home user. 
We generated/collected five different types of noise: (i) White, (ii) Ads, (iii) Music, (iv) News, and (v) Talk shows.  White noise was generated through a random distribution with zeros mean and unique variance. Ads published on the French television between 2018 and 2020 were crawled from the \emph{culturepub.fr} website. Musics were taken from our personal libraries and chosen specifically to be popular songs in France within the same period as used for ads according to \emph{billboard}. 

All noise data were concatenated for each type of noise and then split into three partitions, i.e., training, development, and test, so that noises are unique per partition. The length of each noise type per partition is given in Table \ref{tab:noiseDurations}.

\subsubsection{Adding noises to speech utterances}
We randomly selected a portion of the audio noises with the same duration as the target speech and then added them together, with the noise being multiplied by a gain ($g_n$) following~\cite{eyben2015real}:

\begin{equation}
  g_n = 10^{(\log(g_s) - \frac{SNR}{20})}
  \label{equ:NoiseMixSNR}
\end{equation}
where $g_s$ is the gain of the clean speech file and $SNR$ is Signal-to-Noise Ratio. 

\begin{table*}[t]
\footnotesize
  \caption{Duration of the partitions of the CV corpus used for training, developing, and testing our VAD systems.}
  \label{tab:dataset}
  \centering
  
  \begin{tabulary}{1.0\textwidth}{p{0.4\textwidth}|p{0.1\textwidth}|p{0.1\textwidth}|p{0.1\textwidth}}
        \textbf{Speech type} & \textbf{Training (hh:mm:ss)} & \textbf{Development (hh:mm:ss)} & \textbf{Test (hh:mm:ss)} \\
        \hline
        \hline
        
        Close-talking microphone, crowd-sourced & 2:08:07 & 2:08:04 & 2:09:03 \\ \hline

  \end{tabulary}
\end{table*}
\begin{table*}[t]
\footnotesize
  \caption{Total duration collected for the different types of investigated additive noises.}
  \label{tab:noiseDurations}
  \centering
  
  \begin{tabulary}{1.0\textwidth}{p{0.2\textwidth}|p{0.1\textwidth}|p{0.1\textwidth}|p{0.1\textwidth}|p{0.15\textwidth}}
        \textbf{Partition} & \textbf{Ads} & \textbf{Music} & \textbf{News} & \textbf{Talk shows} \\
        \hline
        \hline
        
        \textbf{Length (hh:mm:ss)} &
        1:38:43 & 1:52:09 & 1:21:04 & 1:21:27 \\ \hline
        
  \end{tabulary}
\end{table*}
\begin{table*}[t]
\footnotesize
  \caption{Summary of the different W2V2 representations of speech used for VAD.}
  \label{tab:SSLs}
  \centering
  
  \begin{tabulary}{1.0\textwidth}{p{0.13\textwidth}|p{0.15\textwidth}|p{0.13\textwidth}|p{0.25\textwidth}|p{0.18\textwidth}}
         \textbf{Name} &  \textbf{\# Training epochs} & \textbf{ Hours of speech} & \textbf{Speech Language} & \textbf{Speech type} \\
        \hline
        

        \textbf{wav2vec2-FR-2.6K-base \cite{evain2021lebenchmark}} & 500k & 2.6k & French & Read \\ \hline
        \textbf{wav2vec2-FR-3K-base \cite{evain2021lebenchmark}} & 500k & 3k & French & Read, Spontaneous, Emotional \\ \hline
        \textbf{wav2vec2-FR-3K-large \cite{evain2021lebenchmark}} & 500k & 3k & French & Read, Spontaneous, Emotional \\ \hline
        \textbf{wav2vec2-large-xlsr-53 \cite{conneau2020unsupervised}} & 250k & 56k & 53 languages & Read \\ \hline
        \textbf{wav2vec2-large-xlsr-53-french \cite{conneau2020unsupervised}} & 250k+20k & 56k+353 & 53 languages + fine-tuning on French & Read \\ \hline

  \end{tabulary}
\end{table*}

\subsection{Representations}
\subsubsection{Mel-scale Filter Bank (MFB)}
As acoustic low-level-descriptors (LLDs), we extracted the first 80 Mel-scale Filter Bank (MFB) coefficients from the audio signal. The feature extraction is realised on a 25\,ms window that is shifted forward in time each 10\,ms. All features were normalised to have zero mean and unit variance according to statistics computed on the training partition, excepted for our investigations on real-time robustness, cf. Section \ref{sec:limitData}, where normalisation is performed for each instance.

\subsubsection{Wav2vec2}
W2V2 is a recent popular SSL model that exploits a contrastive predicting model to extract acoustic representations of speech~\cite{baevski2020wav2vec}. Taking benefits from our previous investigations on SSL representations for downstream speech tasks on French~\cite{evain2021lebenchmark,evain2021task}, we exploited different W2V2 models specifically trained on different types of French speech, as well as a model trained on multilingual data 
~\cite{conneau2020unsupervised}. The list of the different W2V2 models used here is brought in Table \ref{tab:SSLs}. These models, once trained, are supposed to provide speech representations that are more contextualised and less impacted by noises compared to acoustic features obtained with LLDs.

\subsection{VAD model}
\label{sec:Models}

RNN based models, using mainly LSTM and lately GRU, have long been used in speech related tasks in order to reach state-of-the-art results, because of their ability to model the context of data \cite{eyben2013real, wang2019rnn, heitkaemper2020statistical}. Thus, we also experiment with a GRU model, followed by a linear layer, to map the hidden size of the GRU to the number of outputs (here one), and a tangent hyperbolic function, to map the output to the range of $[-1, +1]$, with $-1$ and $+1$ representing non-speech frames and speech frames, respectively.

Also, as fine-tuning has been reported to improve the results of W2V2 models in several different studies in related domains~\cite{evain2021task, conneau2020unsupervised}, here we also tried fine-tuning our W2V2 models for the task of VAD, i.e. we did not freeze any W2V2 parameter while training the VAD. However, we faced convergence issues that require further investigations.

\section{Experiments}

All experiments were performed using the open-source SpeechBrain toolkit~\cite{speechbrain}. We explain in the followings the set of hyper-parameters used for our VAD model, followed by the experiments that were conducted to evaluate its real-time performance. 

\subsection{Hyper-parameters}
In order to save computation time, hyper-parameters were defined only for the normalised MFB features, because they are still the most widely used audio descriptors for speech processing in the literature~\cite{latif2020deep}. For selecting the number of layers and nodes for our GRU model, we used the following range of complexity: 1 layer with 64 nodes (1L64N), 2 layers with 128 nodes (2L128N), and 4 layers with 256 nodes (4L256N). Adam optimiser with 0.01, 0.001, and 0.0001 learning rates was used for training the model. We chose the GRU-1L64N trained with a 0.001 learning rate because it reached very close performance to our best setup, which was GRU-4L256N with 0.0001 learning rate, while requiring much less training time.

\subsection{Robustness for real-time systems on limited data}
\label{sec:limitData}
Current real-time -- or close to real-time -- systems receive packets of data from users over very short spans of time (e.g. 10\,ms) \cite{loreto2014real}. Thus, a VAD system based on deep learning would exhaust resources of a modern system today if it would need to consume every new packet of data received from the user. Therefore, a buffer is defined to continuously store the data, and then, at certain intervals of time (e.g. 1\,s), the VAD model is run only on the data contained in the buffer to detect speech segments. 

As previously explained, a VAD system using normalised MFB features is highly susceptible to the amount of data available for normalisation, which are very limited when using an audio buffer. Hence, we limited in this experiment the size of the audio data when testing our models, to evaluate the impact of a VAD system functioning in (close-to) real-time constraints. This was achieved by cutting each input signal into smaller parts equivalent to an audio buffer, which would later be used as the model's inputs. Thus, in this experiment, we divide the input sequence with different window lengths with respect to the whole available audio input. Using a range of different lengths for the buffer size allows us to evaluate how different speech representations perform when having access to different amounts of input data. We believe this test to be based on a realistic scenario for comparing normalised MFB features with W2V2 representations, when having less access to \textit{a priori} data of a new domain.  

\section{Results}
\label{sec:Res}

\begin{figure*}[t]
\centering
\includegraphics[width=1\linewidth]{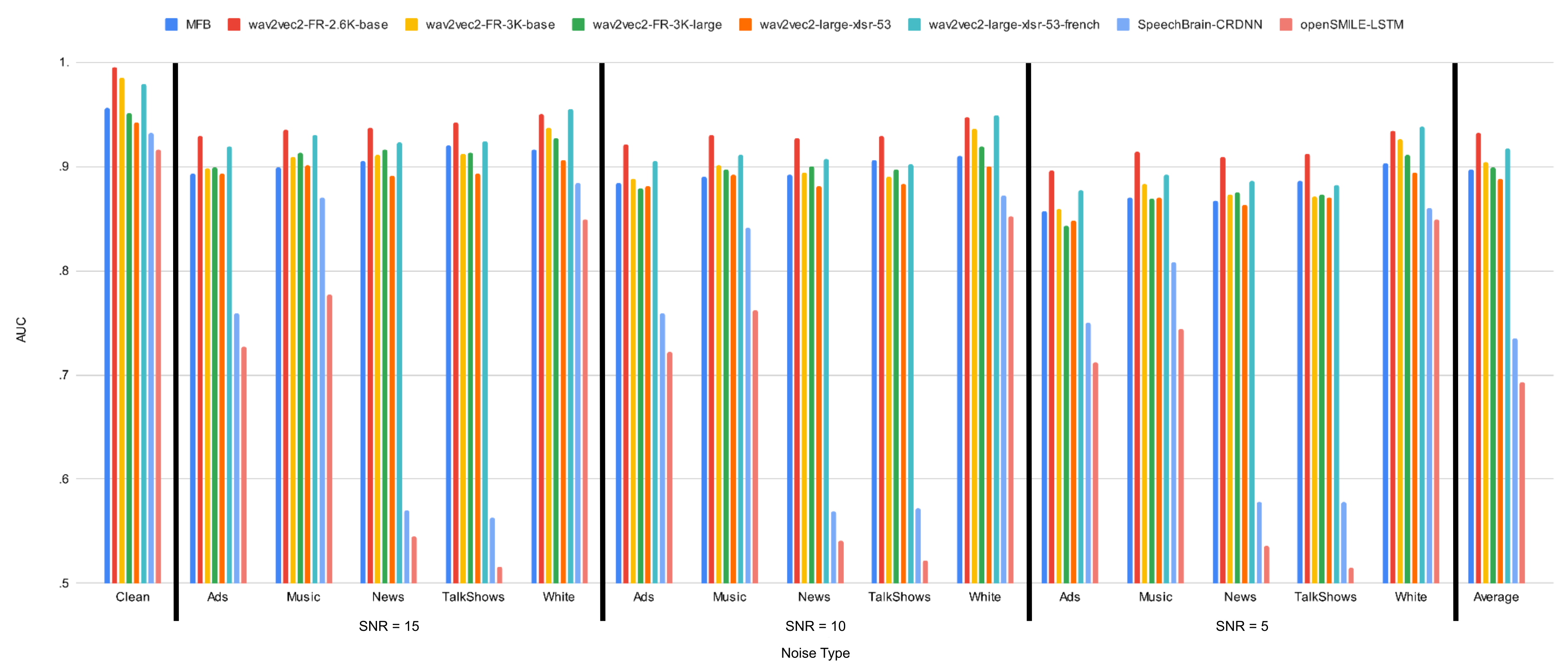}
\caption{Comparison of performance, as measured by the Area Under the Curve (AUC), obtained by our GRU based VAD model using either MFB features or W2V2 representations, for different types of noises (Ads, Music, News, TalkShows and White), and SNRs (5-10-15), on the test set. We also report the performance obtained with two commonly used off-the-shelf VAD systems (SpeechBrain-CRDNN~\cite{speechbrain}, and openSMILE-LSTM~\cite{eyben2013real}).}
\label{fig:W2V2sMFB}
\end{figure*}

Figure \ref{fig:W2V2sMFB} contains different types of comparisons between different representations, noise types, and off-the-shelf models. The metric used for the reported results is Area Under the Curve (AUC) of the Receiver Operating Characteristic (ROC) curve.



\subsection{Comparing Representations}


Even thought our wav2vec2-FR-2.6K-base model significantly outperforms classical MFB features in all settings, reaching almost perfect results on clean speech, MFB features can still achieve comparable results to state-of-the-art.

One may further note that, the performance of W2V2 representations highly depends on the data used for training them. For example, the multilingually trained W2V2 is worse on average than the same model fine-tuned for French data, which shows that VAD models based on W2V2 representations are language dependent. Thus, using contextualised representations, trained for the same language being detected, helps improve the performance.

Furthermore, the W2V2 large models used here, which use a normalisation layer, do not perform better than the smaller W2V2 base models, which do not use any normalisation. Since the best results is achieved by a W2V2 architecture, we can conclude that once trained with enough data from different domains, W2V2 representations can perform much better than classical acoustic features, without the need to be normalised to the new domain using data at hand.

\subsection{Comparison with off-the-shelf VADs}
Here, to compare our models to off-the-shelf VADs, we chose the commonly used open-source OpenSMILE toolkit\footnote{https://www.audeering.com/research/opensmile/}~\cite{eyben2010opensmile} as well as the already available recipe for VAD in the open-source SpeechBrain toolkit\footnote{https://huggingface.co/speechbrain/vad-crdnn-libriparty}~\cite{speechbrain}.



As can be seen from Figure \ref{fig:W2V2sMFB}, all of our models consistently and significantly outperform off-the-shelf VADs. The difference in performance is more highlighted in the presence of noise, suggesting that our scheme of training a VAD model under different noisy scenarios is especially effective for detecting French speech from close-talking microphones. 
We also used the RECOLA dataset, which consists of close-talking French speech, and observed the consistent better performance of our GRU model with wav2vec2-FR-2.6K-base representation compared to off-the-shelf VADs still holds true; on average, 0.921 AUC for our system vs 0.877 AUC for SpeechBrain-CRDNN and 0.804 for openSMILE-LSTM.

\begin{figure}[t]
\centering
\includegraphics[width=.75\linewidth]{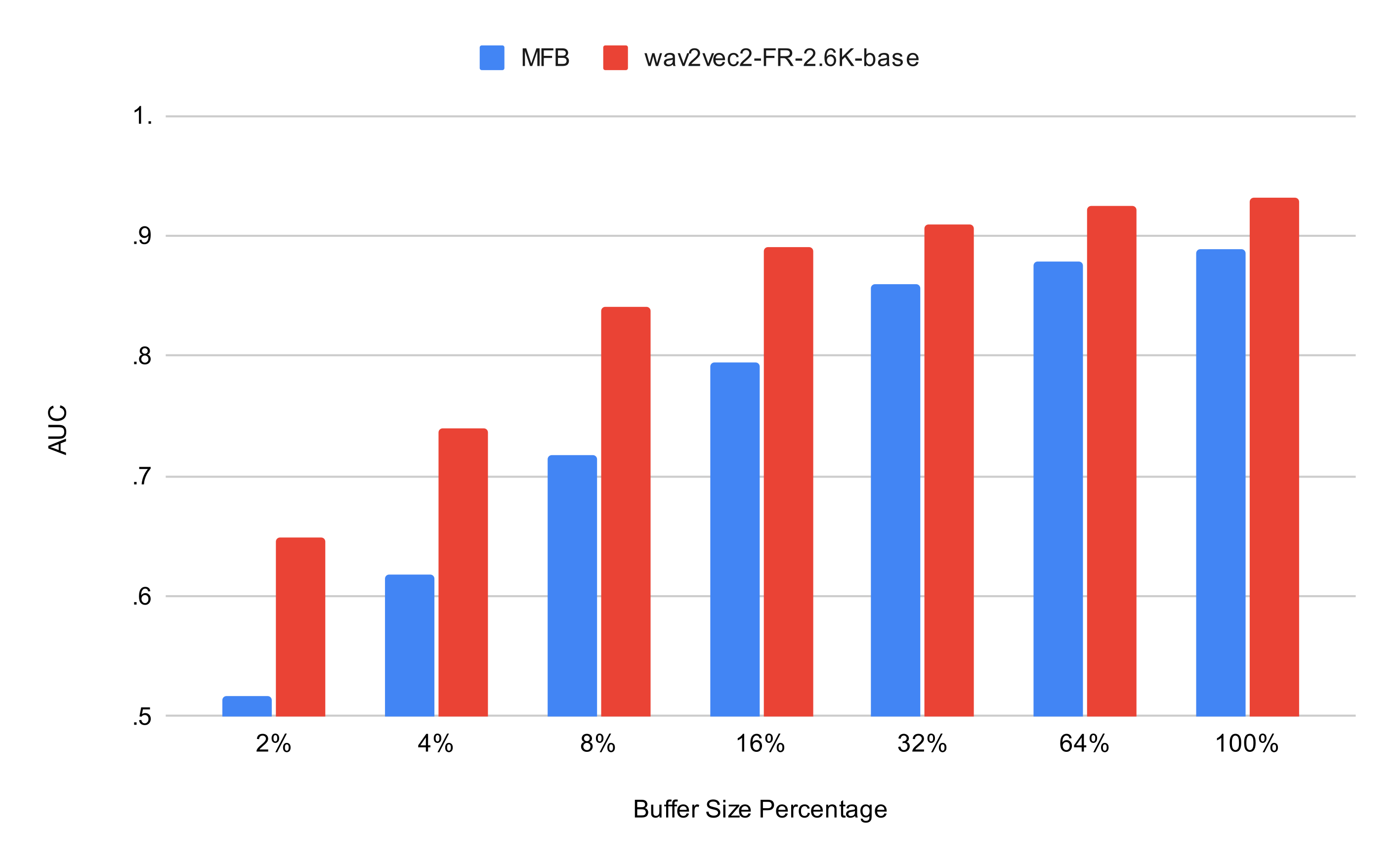}
\caption{The effect of changing the input buffer size on the results of the GRU based VAD system, using normalised MFB features and the wav2vec2-FR-2.6K-base representations. 
}
\label{fig:buffer}
\end{figure}

\subsection{Limiting the buffer size for real-time analysis}

For this experiment, we picked wav2vec2-FR-2.6K-base representation, since it achieved the most performant VAD, and compared it to normalised MFB features. We considered a range of different lengths for our buffer to compare the two representations (c.f. Section \ref{sec:limitData}). Figure \ref{fig:buffer} shows the performance of the model considering different buffer lengths, and thus different amounts of data as input. Since our samples have at least one minute length, and can be some seconds longer depending on the speech utterance, the 100\% represents at least one minute in length. Thus, we can see that while the performance of MFB features are very poor when faced with 4\% of data (around 2.5\,s), the W2V2 representations can perform well. Using only 16\% of the buffer size (around 10\,s), we can see that the W2V2 representation can already pass the best performance of the normalised MFB features, with all the data available to it. This clearly shows that W2V2 representations with no normalisation can outperform traditionally used normalised MFB features, with less need for data from the target domain.


\section{Conclusions}
\label{sec:conclusion}
In this paper, we focused on how SSL methods, especially W2V2 representations, can help VAD systems to achieve robust predictions on French speech recorded by close-talking microphones at home. We showed that such representations trained for the target language can help the VAD to perform better than traditional methods across different noise types. We also showed that W2V2 representations need less \textit{a priori} information, in order to obtain better results than normalised MFB features commonly used today. 

However, one issue with W2V2 representations compared to MFB features is that they are slower and require more computational resources. Thus, we realised a system that first uses classical MFB based VAD to first detect possible segments, and then activate a more accurate W2V2 based VAD to verify and more accurately detect the speech segments. We were able to run this system on a normal CPU (1.4 GHz Quad-Core Intel Core i5) in real-time (under 1\,s) for a buffer size of 15\,s.


\section*{Acknowledgment}
The research leading to these results has received funding from the Association Nationale de la Recherche et de la Technologie (ANRT), under grant agreements No. 2019/0729 (Wellbot project).

\bibliographystyle{unsrtnat}
\bibliography{references}  

\end{document}